\documentclass{aa}  

\usepackage{graphicx}
\usepackage{txfonts}
\usepackage{xcolor}
\begin{document}

   \title{Hierarchical star formation in nearby galaxies\thanks{Table 1 is only available in electronic form
at the CDS via anonymous ftp to cdsarc.u-strasbg.fr (130.79.128.5)
or via http://cdsweb.u-strasbg.fr/cgi-bin/qcat?J/A+A/}}

\author{
M. J. Rodríguez\inst{1}\thanks{jimenaro@fcaglp.unlp.edu.ar},
G. Baume\inst{1,2} and 
C. Feinstein\inst{1,2}
}
\institute{
Instituto de Astrofísica de La Plata (CONICET-UNLP), Paseo del bosque S/N, 
La Plata (B1900FWA), Argentina, \and 
Facultad de Ciencias Astronómicas y Geofísicas - Universidad Nacional de La Plata, Paseo del bosque S/N, La Plata (B1900FWA), Argentina}

\date{Accepted XXX. Received YYY; in original form ZZZ}

 
  \abstract
   {}
   {The purpose of this work is to study the properties of the spatial distribution of the young population in three nearby galaxies in order to better understand the first stages of star formation.}
   {We used ACS/HST photometry and the "path-linkage criterion" in order to obtain a catalog of young stellar groups (YSGs) in the galaxy NGC~2403. We studied the internal distribution of stars in these YSGs using the $Q$ parameter. We extended these analyses to the YSGs detected in in NGC~300 and NGC~253 our previous works. We built the young stars' density maps for these three galaxies. Through these maps, we were able to identify and study young stellar structures on larger scales.}
   {We found 573 YSGs in the galaxy NGC~2403, for which we derived their individual sizes, densities, luminosity function, and other fundamental characteristics.  
    We find that the vast majority of the YSGs in NGC~2403, NGC~300 and NGC~253 present inner clumpings, following the same hierarchical behavior that we observed in the young stellar structures on larger scales in these galaxies. We derived values of the fractal dimension for these structures between $\sim$ 1.5 and 1.6. These values are very similar to those obtained in other star forming galaxies and in the interstellar medium, suggesting that the star formation process is regulated by supersonic turbulence.
}
   {}


   \maketitle
%

\section{Introduction}

It is generally accepted that the conversion of gas into stars is produced by the gravitational collapse of gas and dust inside the molecular clouds  \citep{2000prpl.conf..151C,2003ARA&A..41...57L,2003RPPh...66.1651L}. However, this is only a general picture.  
The star formation process is not at all well understood; the complexity and the amount of physical mechanisms involved make this an unclear issue  \citep{1993ApJ...419L..29E,2003RPPh...66.1651L,2007ARA&A..45..565M}. 
Moreover, the formation of stars seems to proceed in groups like stellar clusters or associations \citep{2003ARA&A..41...57L}. 
These groups are formed in the densest part of the giant molecular clouds  \citep[GMCs;][]{2020SSRv..216...64K}.
Thus, GMCs present the initial conditions for star and cluster formation.

The interstellar medium (ISM) presents a hierarchical and self-similar distribution that is extended over several space scales \citep[e.g.,][]{1993ApJ...419L..29E,1996ApJ...471..816E}. This means that the large structures in the ISM, like the GMCs, are composed of smaller and denser structures, which in turn split themselves into even smaller and denser structures, and so on. 
The main mechanisms associated with this spatial distribution are turbulence, magnetic fields, and self-gravity \citep{1996ApJ...466..802E}, but again these are still not well understood. 

On the other hand, this hierarchical organization is also observed in the young stellar population \citep[e.g.,][]{2015MNRAS.452.3508G,2017MNRAS.468..509G,2017ApJ...849..149S,2017ApJ...835..171S,2018ApJ...858...31S,2017ApJ...842...25G,2016A&A...594A..34R,2018MNRAS.479..961R,2019A&A...626A..35R}. The spiral arms are composed of several stellar complexes reaching kpc scale, which are formed by several OB associations and other stellar aggregates that do not have a precise name, because when we observe the big picture, we see the young stellar distribution as a continuum, with more concentrated regions that have inner overdensities \citep[e.g.,][]{2018PASP..130g2001G}. The OB associations at the same time present inner open clusters or other subgroups \citep{1964ARA&A...2..213B}. The hierarchical distribution does not seem to end with the clusters, since the distribution of stars inside them can be inhomogeneous and present clumpings \citep{2009ApJ...696.2086S,2012A&A...541A..95F}.

In this context, it is believed that the distribution of stars is inherited from the self-similar behavior of the ISM \cite[e.g.,][]{1999AJ....117..764E,2017MNRAS.467.1313V}; after formation, the stars would evolve to a homogeneous or radial distribution until the structure dissolves in the field  \citep{2006A&A...449..151S}. Although several works present results pointing in that direction, there are very young clusters with homogeneous distribution and old groups with inner substructures \citep{2008MNRAS.383..375C,2009ApJ...696.2086S}. 

The self-similar behavior described above is a property of geometrical objects called fractals. Fractals were introduced by \citep{mandelbrot1982fractal} as objects that have approximately the same shape on all scales. This means that if the object is inspected on very small scales, we see a reduced copy of the original shape.  
The ISM and the distribution of young stars are not evidently fractals in the mathematical sense, because they do not have structures on arbitrarily small scales. However, we can say that in a general sense they have a fractal structure.  

\begin{figure*}
\centering
\begin{tabular}{cc}
\includegraphics[trim={80 30 50 0},width=0.4\textwidth]{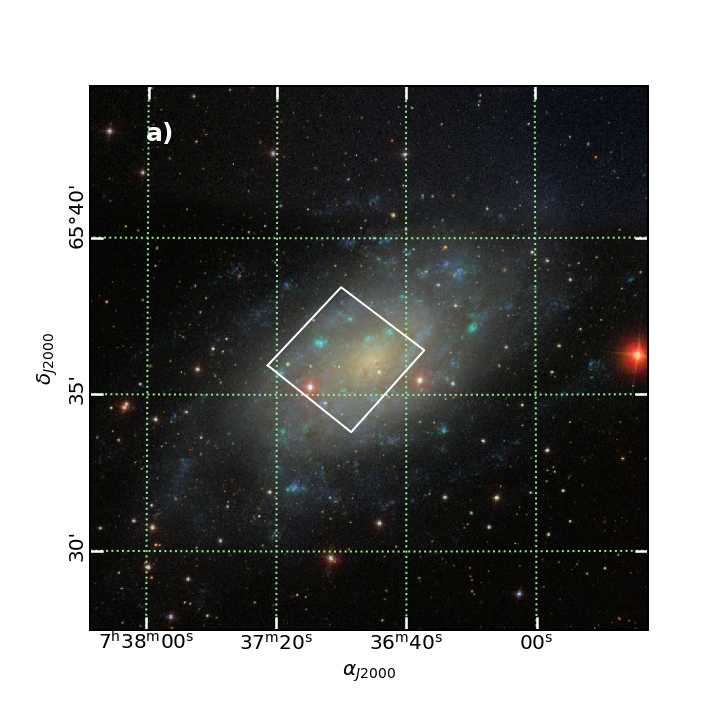} &
\includegraphics[trim={100 30 30 0},width=0.4\textwidth]{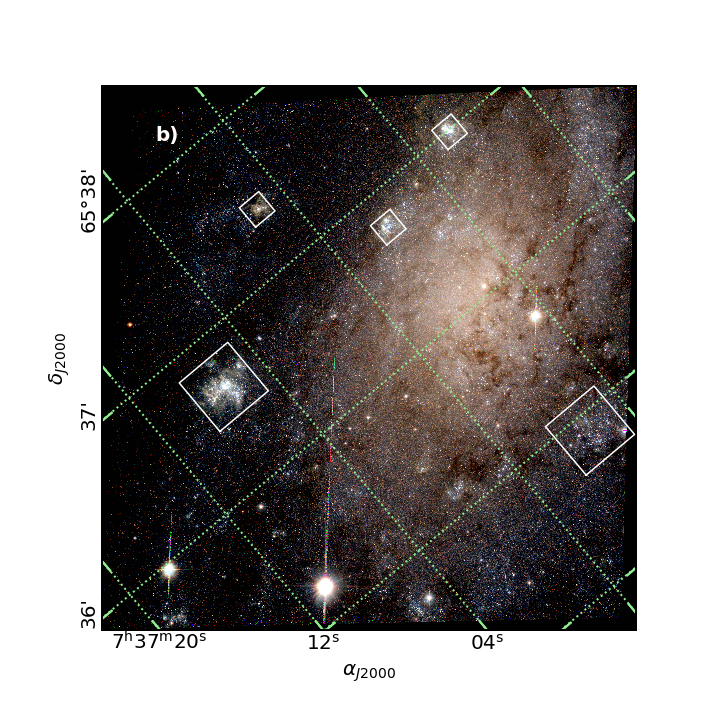}
\end{tabular}
\caption {{\bf a)} SDSS color image of the galaxy NGC~2403 from Aladin, the ACS/WFC observed field is indicated in white. {\bf b)} RGB combined image (B:~$F475W$, G:~$F606W$, and R:~$F814W$) from ACS/WFC data. White squares indicate selected regions with high stellar densities.}
\label{field_acs}
\end{figure*}

\begin{figure}
\includegraphics[trim={40 20 30 50},width=\columnwidth]{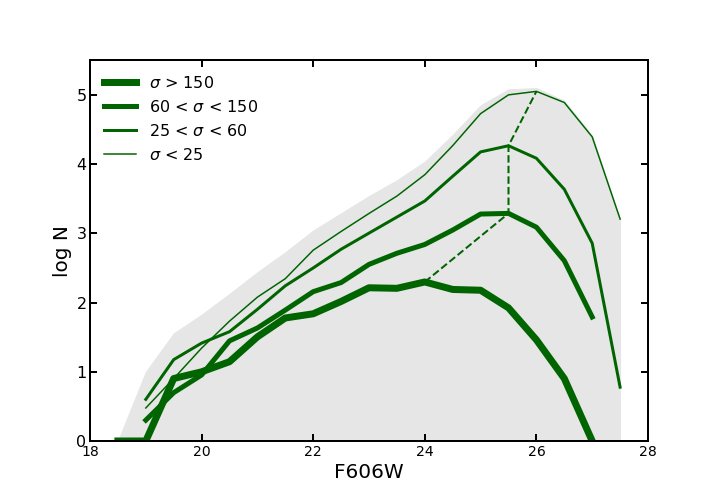}
\caption{LF from NGC~2403 ACS data for regions with different star densities ($\sigma$) of bright stars, indicated in stars per arcsec$^2$. Completeness value in each case is indicated by the slashed line, and it ranges form 24 to 26. Gray area indicates LF for the complete field of view (FOV).}
\label{LF}
\end{figure}

\begin{figure*}
\includegraphics[trim={0 10 0 30},width=\textwidth]{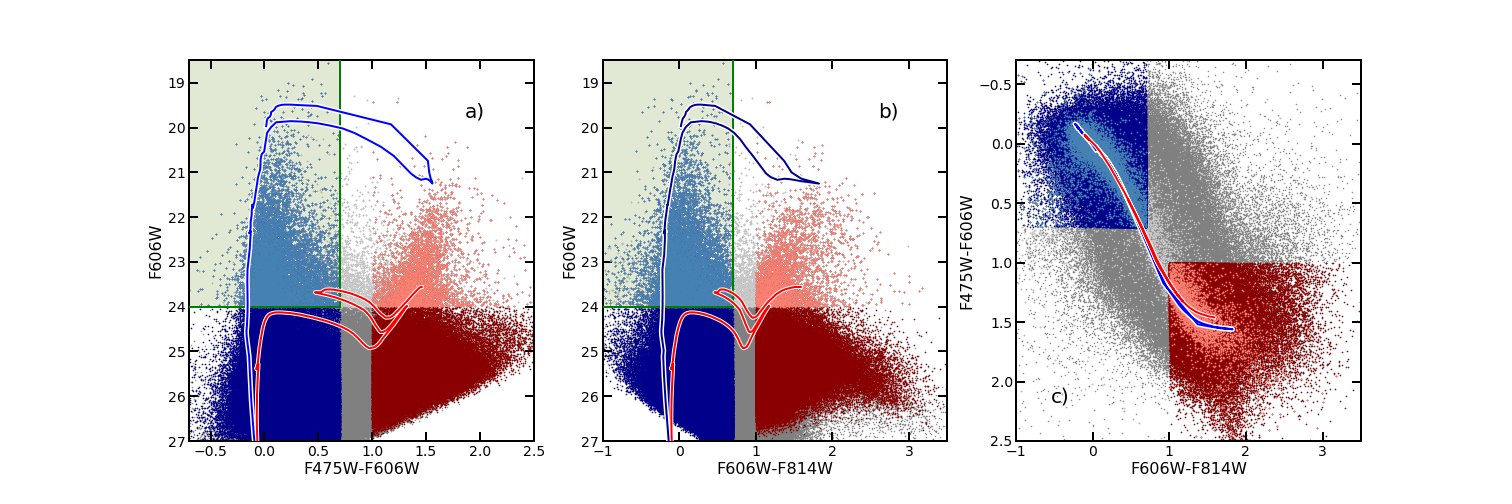}
\caption{Photometric diagrams from ACS/WFC data. Blue and red indicate adopted blue and red objects, respectively, Light colors indicate the adopted bright ones. Green boxes indicate the considered stars in the PLC method. Blue and red curves are the Parsec 1.2S isochrones corresponding to 10 and 90 Myr, respectively (see text for details).}
\label{CMD}
\end{figure*}

In previous works \citep{2016A&A...594A..34R,2018MNRAS.479..961R,2019A&A...626A..35R}, we have searched for and analyzed young star groups in the galaxies NGC~300, NGC~253, and NGC~247. In particular, we observed the hierarchical organization of the young populations in these galaxies. We observed young structures on a wide range of scales, from young stellar groups like clusters or associations, to large stellar complexes, and we traced all the links between them by means of tree diagrams.  
In this work, we used the information obtained in previous works to obtain fractal parameters for NGC~300, NGC~253, and a new galaxy, NGC~2403. For this last galaxy, we present a complete study of their young stellar populations. We compared the fractal values obtained for these three galaxies with those obtained for NGC~247.

The present work is organized as follows. In Sect.~\ref{NGC2403}, we present the search for and analysis of young stellar groups (YSGs) in NGC~2403. In Sect.~\ref{MSTandQ}, we study the internal distribution of young stars inside the YSGs for all the YSGs detected in NGC~2403, NGC~300, and NGC~253. In Sect.~\ref{sect_fractal}, we analyze the young stellar structures on larger scales in the three galaxies mentioned above. In Sect.~\ref{discuss}, we discuss our main results. Finally, in Sect.~\ref{conclusions}, we summarize the findings of this study. 

\begin{table*}[]
\centering
\caption{Catalog of young stellar groups in NGC~2403. Here we present the first ten lines, the complete version is available at CDS.}
\begin{tabular}{ccccccccccccc}
\hline\hline    
Name  & $\alpha_{2000}$ &  $\delta_{2000}$ & $r~["]$  & $N$   & $N_B$  & $F606W_{min}$   & $\Gamma_{LF}$ &  $e_{\Gamma}$ & d$_{GG}~[Kpc]$ & $\rho~[obj./pc^3]$ & $Q$ & $e_Q$\\ 
\hline
AS001 & 114.278295 & 65.610044 & 0.9  & 38 & 17 & 22.34 &  - &  - & 1.75 & 0.00457 & 0.69 &      0.068\\
AS002 & 114.278241 & 65.608591 & 0.88 & 41 & 24 & 21.65 & - & - & 1.79 & 0.00527 & 0.73   & 0.062 \\
AS003 & 114.277129 & 65.613554 & 0.8  & 32 & 9 & 21.77 & - & - & 1.65 & 0.00547 & 0.63 & 0.082\\
AS004 & 114.189149 & 65.616678 & 2.78 & 321 & 177 & 18.56 & 0.35 & 0.02 & 2.0 & 0.00131 & 0.74 & 0.020\\
AS005 & 114.227903 & 65.590895 & 0.54 & 12 & 9 & 21.59  & - & - & 1.47 & 0.00668 & 0.57 & 0.065\\
AS006 & 114.277486 & 65.611102 & 0.72 & 17 & 13 & 20.29 & - & - & 1.7 & 0.00399 & 0.53  &       0.051\\
AS007 & 114.279864 & 65.610909 & 1.32 & 74 & 35 & 20.5  & 0.18 & 0.04 & 1.77 & 0.00282 & 0.59 &      0.046\\
AS008 & 114.214663 & 65.581727 & 0.7  & 27 & 16 & 19.15 & - & - & 2.08 & 0.0069 & 0.77 & 0.074\\
AS009 & 114.271873 & 65.611442 & 0.72 & 30 & 22 & 21.52 & - & - & 1.53 & 0.00704 & 0.57 &        0.060\\
AS010 & 114.204487 & 65.595066 & 0.66 & 19 & 13 & 21.94 & - & - & 0.61 & 0.00579 & 0.66 & 0.069\\
\hline
\end{tabular}
\begin{minipage}{0.9\linewidth}
{\bf Columns:} 
(1): ID; (2) \& (3): coordinates; (4): radius in arcsec; (5): numbers of stars in the YSG; (6): number of stars members with $F606W <$24; (7): $F606W$ value of the brightest blue star; (8): luminosity function slope; (9): error in the LF slope; (10): galactocentric distance in Kpc; (11): stellar density; (12): $Q$ value (see Sect.~\ref{MSTandQ}); 
(13): error in $Q$.
\end{minipage}
\label{groups_catalog}
\end{table*}

\begin{figure*}
\includegraphics[trim={80 50 80 70},width=\textwidth]{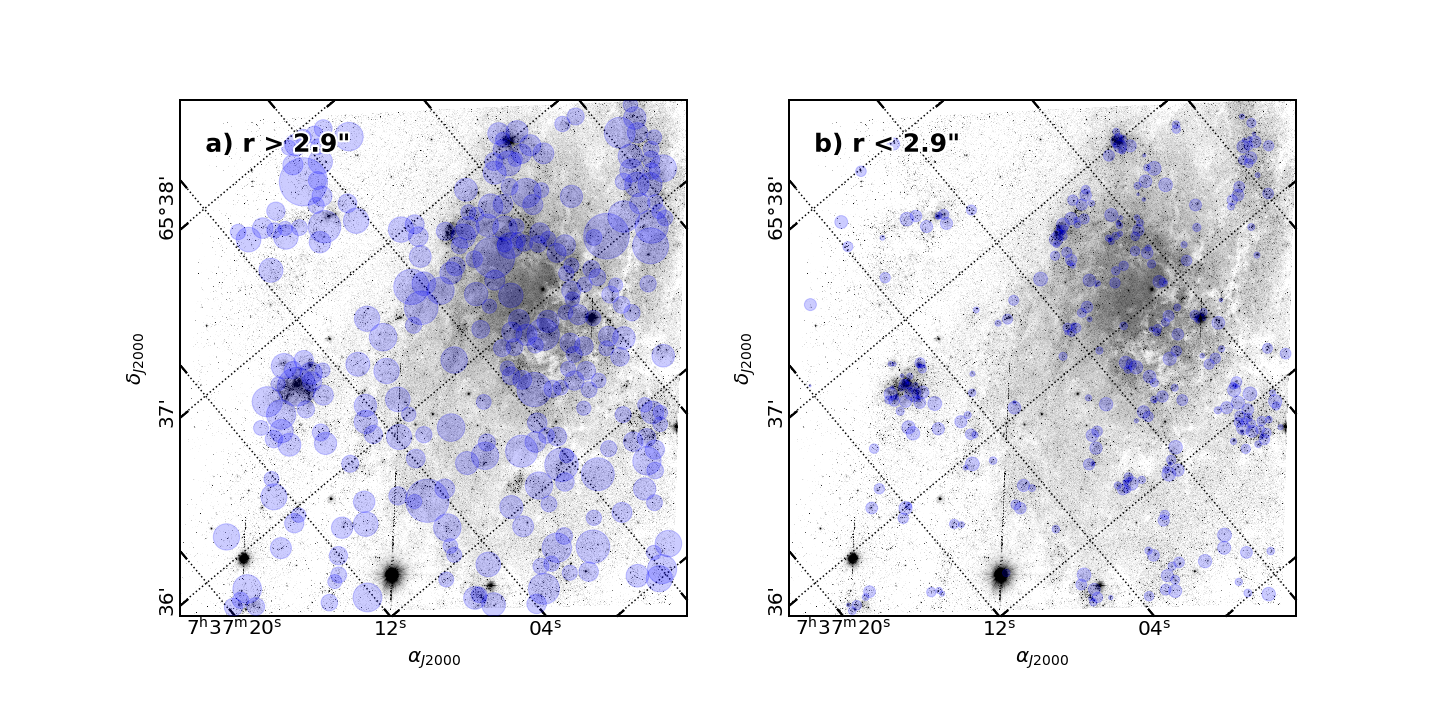}
\caption{Finding charts showing spatial distribution of identified YSGs. For clarity, two panels are presented with larger and smaller YSGs than the mean radius.}
\label{fig:chart3}
\end{figure*}

\section{Young stellar groups in NGC~2403}
\label{NGC2403}

The SAcd~C galaxy NGC~2403 \citep{2015ApJS..217...27A} is a member of the M~81 group. It is located at 3.18 Mpc \citep{2013AJ....146...86T}, which corresponds to projected linear scale of $\sim$ 15.4 $pc/"$.
This galaxy is rich in HII regions, star forming
complexes, and stellar clusters \citep[e.g.,][]{1985PASP...97.1065H,1990A&A...237...23S,1999AJ....117.1249D}.
Several of its HII regions are giant, some of them as bright as the 30 Doradus complex in the Large Magellanic Cloud (LMC), which is the most massive starburst region in the Local Group. The large number of star forming regions that this galaxy presents, added to its proximity, make NGC 2403 an ideal target for our study.

\subsection{Data} 
\label{data} 

We used images and photometric data of one central field of the galaxy (see Fig.~\ref{field_acs}). These data correspond to The ACS Nearby Galaxy Survey (ANGST) and are available in the HST~MAST archive\footnote{https://archive.stsci.edu} \citep[see][for data reduction and photometry details]{2009ApJS..183...67D}. The photometric tables provide information on the filters $F475W$, $F606W$, and $F814W,$ and the observations were carried out in August 2004 during the HST cycle 13 as part of the GO-10182 program (PI: A. Filippenko). Only data with $chi < 15$ were considered to avoid spurious detections, mainly due to bright foreground saturated stars. Exceptions to this rule were some particular regions with high stellar densities, as indicated in Fig.~\ref{field_acs}b.

To evaluate the completeness of the sample, we built the $F606W$ luminosity function (LF) for regions with different bright star densities ($F606W < 24$), and these are presented in Fig.~\ref{LF}. This figure reveals that the number of stars begins to decrease for $F606W$ values in the range from 24 to 26, depending on the region crowding.

Over the color magnitude diagrams (CMDs; Fig.~\ref{CMD}), we divided our sample between blue stars ($F475W - F606W < 0.7;F606W - F814W < 0.7$) and red stars ($F475W - F606W > 1.0$ ; $F606W - F814W > 1.0$). Then, we selected the bright blue stars: these are the blue stars with magnitudes $F606W < 24$, and they correspond to the upper main-sequence population. These stars fall inside the green rectangle in Fig.~\ref{CMD}. 

We compared this selection with evolutionary models of Parsec version 1.2S \citep {2014MNRAS.445.4287T} with solar metallicity $Z_{\odot}$=0.0152 in order to estimate a range of ages for this population. The blue and red isochrones in Fig.~\ref{CMD} correspond to 10 and 90 Myr, respectively. So, we assumed an age range for the selected young population between these values. For the displacement of the evolutionary models, we adopted a distance modulus of 27.51, a normal reddening law ($R = A_{V} /E(B-V) = 3.1$), and a value of $E(B-V)$ = 0.034 corresponding to the foreground reddening toward NGC~2403 \citep{2011ApJ...737..103S}. The selected red population is composed of red He-burning stars, the asymptotic giant branch (AGB), stars and the red giant branch (RGB) stars.

\subsection{Identification and analysis of young stellar groups}
\label{PLC}

For the identification and analysis of the YSGs in NCG~2403, we used the methods described in our previous work, \cite{2016A&A...594A..34R}, and we advise the reader to refer to this paper for details about the procedure. We only summarize general aspects relevant for the current analysis here.

We used the path linkage criterion (PLC; \citealt{1991A&A...244...69B}) to search for YSGs. This is a friend-of-friend algorithm based on the distance between stars. We applied the PLC to the stars belonging to the blue bright sample with a search radius ($d_s$)from 0.3-2$"$ and a minimum number of stars $p$=8. These values were chosen observing the distribution of stellar groups detected as a function of $d_s$ for different values of $p$ \citep{1991A&A...244...69B}. We also selected a range of values for $d_s$, instead of a single value, because it is important not to miss the smallest or largest groups \citep{2016A&A...594A..34R, 2018MNRAS.479..961R, 2019A&A...626A..35R}. We found 573 YSGs, these groups are mainly OB associations, but as the method works by linking stars by distance, it is probable that some large open clusters or stellar aggregates larger than a typical association have been detected, so YSG refers to all the structures detected by the PLC.

We computed the fraction of detected YSGs that could be stochastic
fluctuations of field stars. In order to do that, we ran numerical simulations with random 
distribution of stars and similar densities to the densest part of the observed field.
Then we applied the PLC method over them. We described this procedure in \cite{2018MNRAS.479..961R}. We estimated a typical surface density over the spiral arms of $\Sigma = 0.517 stars~arcsec^{-2 }$. Using this value, we obtained only five YSGs when we ran the PLC with a search radius of  $d_s = 0.4"$ and 247 YSGs with $d_s = 0.7"$ in a total of 10,000 experiments. These fractions are very small, so most of the YSGs detected should be real ones.
   
We studied the YSGs in a systematic and homogeneous way by means of our analysis code \citep{2016A&A...594A..34R, 2018MNRAS.479..961R}. We estimated sizes, decontaminated CMDs, LFs, numbers of members, densities, and galactocentric distances for each one of the 573 YSGs. In particular, we computed the LF slopes for their brightest regions ($F606W < 24$) and only took into account those YSGs with enough stars to obtain reliable results, meaning YSGs with at least 40 stars and four bins with a nonzero number of stars. As a result, we built a catalog of these properties. In Table~\ref{groups_catalog}, we show the first ten lines, the completed version is available online at the CDS. Their spatial distribution is presented in Fig.~\ref{fig:chart3}, where small and large YSGs were separated. We notice that the large ones (panel a) better delineate the galactic spiral structure and they revealed the presence of a ring surrounding the galactic center. This latter structure would be $\sim$ 15" in radius ($\sim$ 240~pc).

\begin{figure}
\includegraphics[trim={0 25 0 20},width=\columnwidth]{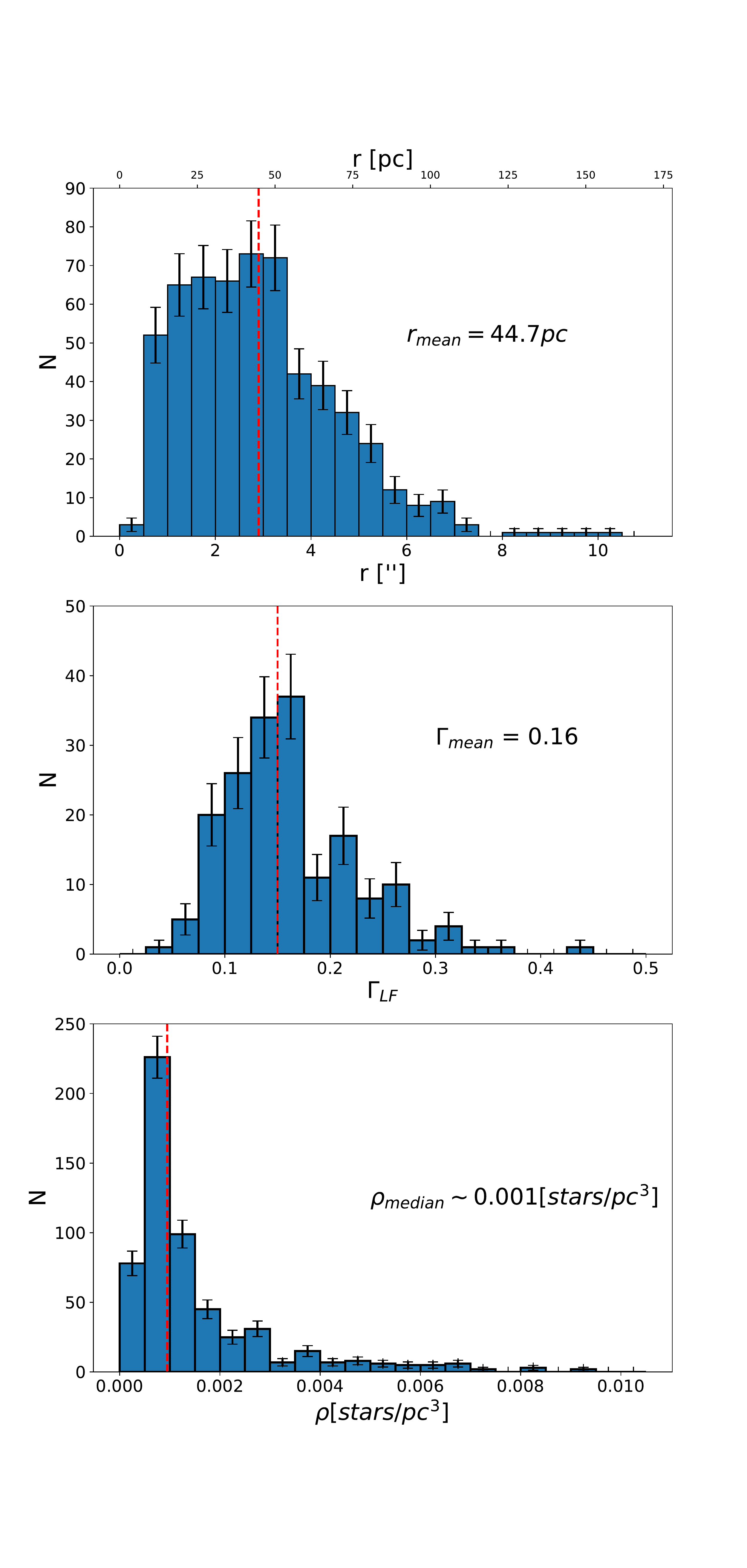}
\caption{Distribution of sizes (top), LF slopes (middle), and densities (bottom) for identified YSGs in NGC~2403.}
\label{histograms}
\end{figure}

In Fig.~\ref{histograms}, we show the distribution of sizes, LF slope ($\Gamma$), and densities for the PLC groups. The size distribution shows a wide range of YSG radii, from 7~pc to 155~pc. We also obtained a mean value of $\sim$45~pc and a median of $\sim$42 pc. We obtained very similar values in our previous studies over other nearby galaxies (see Table~\ref{tab:parameters}). In particular, for NGC~253 \citep{2018MNRAS.479..961R} we computed mean and median values of 47~pc and 40~pc, respectively, whereas in NGC~300 we found typical values of 50~pc for the YSGs \citep{2016A&A...594A..34R,2018MNRAS.479..961R}. The YSGs in NGC~247 presented slightly larger values around $\sim$60~pc, reaching sizes up to 200~pc. Thus, NGC~247 presents the largest YSGs among the four galaxies. 

\begin{figure*}
    \centering
    \resizebox{\hsize}{!}{\includegraphics{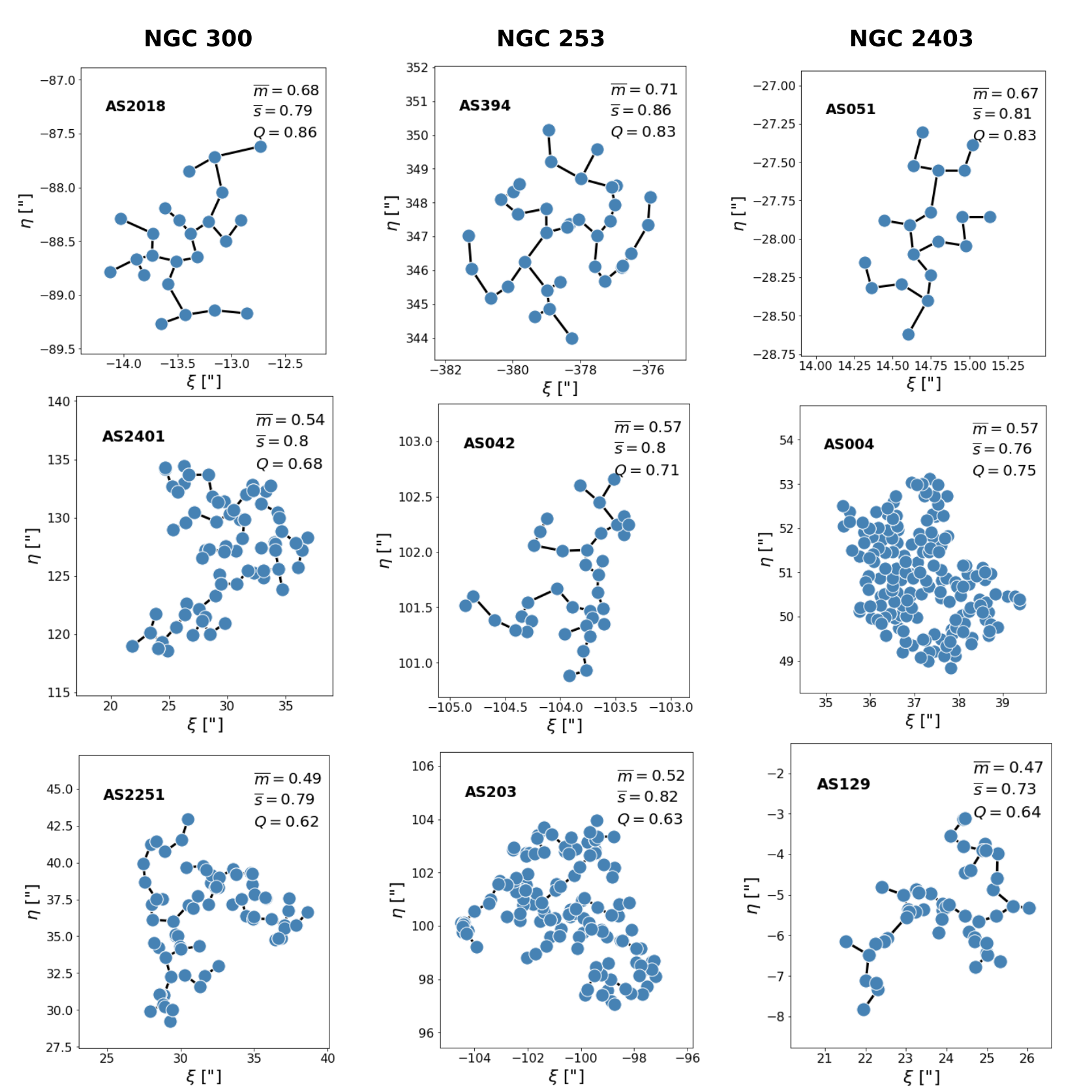}}
    \caption{MST for three YSGs in each galaxy. The positions of the stars are indicated with blue circles and black lines represent the corresponding MST. The first column corresponds to YSGs in NGC~300, the second one to NGC~253, and the last column to NGC~2403. In each case, we show MST with different values of $Q$. The $\overline{m}$, $\overline{s}$ and $Q$ values are also indicated in the upper-right corner of each panel.}
    \label{mst}
\end{figure*}

\begin{figure*}
\centering
\resizebox{\hsize}{!}{\includegraphics{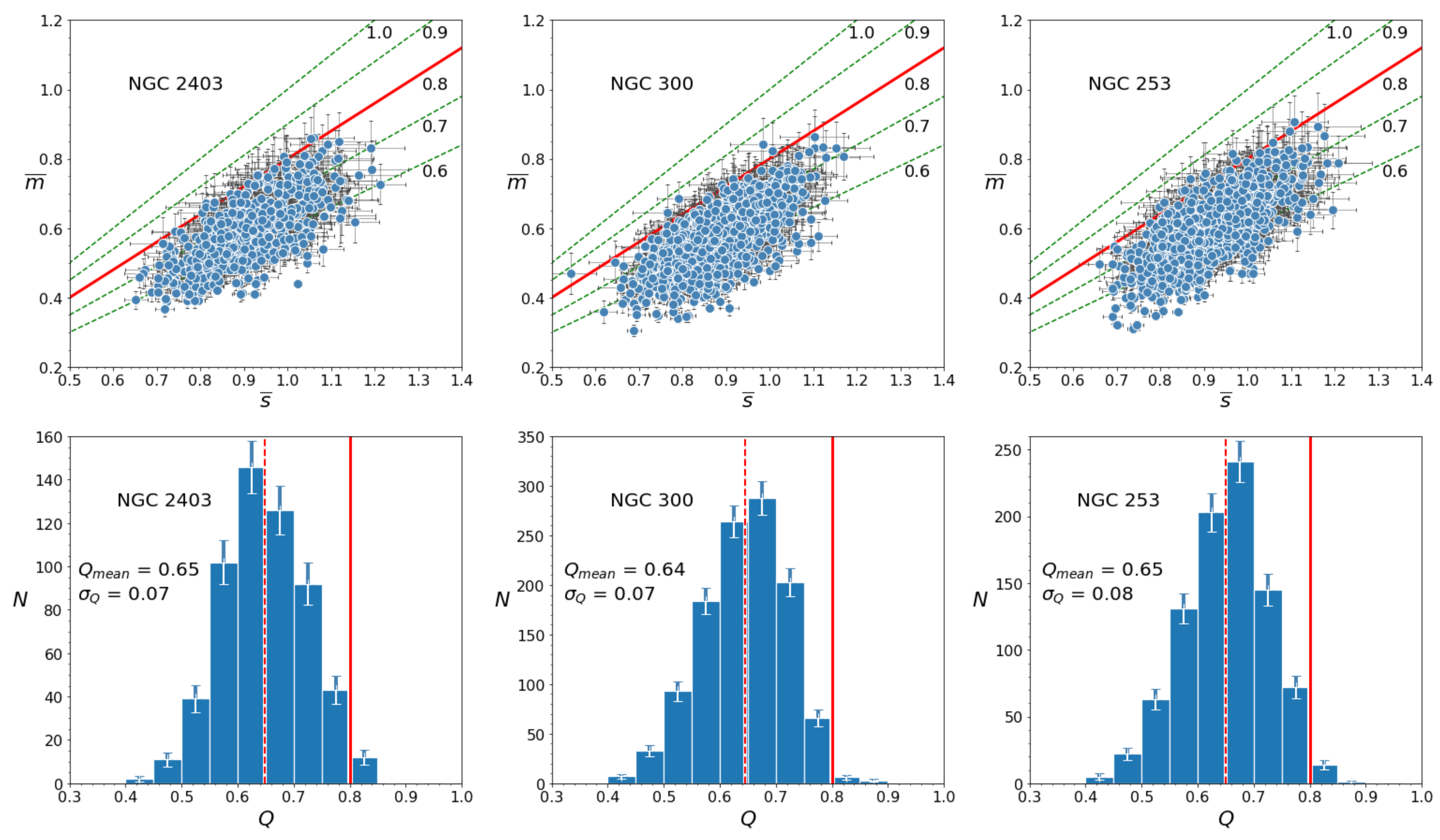}}
\caption{Top panels show $\overline{m}$ vs. $\overline{s}$ for the YSGs located in each galaxy. The red continuous line indicates $Q = 0.8$, the dashed green lines indicate other Q values. The bottom panels show the $Q$ distributions for the three galaxies. We indicate the mean $Q$ value and its standard deviation. The mean value is also indicated with a red dashed line, the red continuous line corresponds to $Q = 0.8$. The error bars were computed as $\sqrt N$.}
\label{FigQ}
\end{figure*}

Regarding the LF slopes, YSGs in NGC~2403 present values (Fig.~\ref{histograms}, middle panel) from 0.04 to 0.43, and mean value of $\Gamma$ = 0.16. The mean value obtained for this parameter in other galaxies are similar and also presented in Table~\ref{tab:parameters}. In particular, for NGC~247 we obtained a bimodal shape of the LF slope distribution with peaks at  $\Gamma \sim$ 0.1 and $\sim$ 0.2.

In the bottom panel of Fig.~\ref{histograms}, we show the density distribution. According to the sample completeness (see Fig. \ref{LF}), we only considered stars with M$_{F606W}$ $\lesssim$~-0.54 that correspond to $B7$ spectral type. We observed a peak in the group densities of $\sim$ 0.001 $stars.pc^{-3}$. 

\section{The internal structure of the young stellar groups}
\label{MSTandQ}

\subsection{Methodology}

We studied the internal distribution of stars inside the YSGs, in order to see if they present a homogeneous distribution or they have inner sub-clumpings. In this last scenario, the distribution of stars still follows the hierarchy derived from the molecular cloud fragmentation. 

To study this distribution, we used the $Q$ parameter, which was introduced by \citet{2004MNRAS.348..589C} to decide if a distribution of points is fractal or not. We used this method in \cite{2019A&A...626A..35R} for the YSGs in NGC~247. In the present work, we extend the analysis to NGC~2403, NGC~253, and NGC~300. For these last two galaxies, we had obtained 875 and 1147 YSGs, respectively \citep{2016A&A...594A..34R,2018MNRAS.479..961R}.

We first built the minimum spanning tree (MST) for all the YSGs. The MST is the shortest network of straight lines that connect all the stars in the group without closed loops. The distances between two connected points are called edge length $(m)$ and $s$ is the separation between any two stars. The $Q$ parameter is defined as the ratio between the mean edge length normalized by the cluster area and the mean separation of all the stars, normalized by the cluster ratio. That is, $Q$=$\overline{m}$ / $\overline{s}$, where \\
$\displaystyle\overline{m}=\frac{1}{(AN)^{1/2}}$$\displaystyle\sum_{i=1}^{N-1}m_i$\\
$\displaystyle\bar{s}=\frac{2}{N(N-1)R_{sc}}\displaystyle\sum_{i=1}^{N-1}\displaystyle\sum_{j=i+1}^{N}\left|\vec{r_i}-\vec{r_j} \right|$.\\
Here, $N$ is the number of stars, $A$ is the area of the smallest circle that contains all the stars, R$_{sc}$ is its corresponding radius and $r_{i}$ is the position of the i star. 
The error in the value of $Q$ was estimated using the bootstrap method \citep{2015MNRAS.448.2504G,2019A&A...626A..35R}.

\cite{2004MNRAS.348..589C} found that $Q \sim 0.8$ corresponds to groups with a uniform distribution, while groups with $Q$ increasing from 0.8 to 1.5 present a more central concentrate distribution. On the other hand, groups with values of $Q$ decreasing from 0.8 to 0.45 present substructures.

\subsection{Results}

The obtained $Q$ values for each one of the 573 YSGs and their corresponding error are listed in Table~\ref{groups_catalog}.
We obtained that most YSGs have $Q$ $\textless$ 0.8, this means that they present sub-clumpings. We only found very few YSGs with values greater than 0.8, but as we can see in Fig.~\ref{FigQ}, these values are just above 0.8 and could be explained by the uncertainty. Furthermore, these YSGs in general have few stars, just the minimum number that we considered in the PLC or a few more, so the uncertainty for $Q$ should be bigger.
The percentages of YSGs with $Q$ values above 0.8 are very small: only 0.8\% in NGC~300, 1.67\% in NGC~253, and 2\% in NGC~2403. 

The disk of the galaxies where the YSGs are located could have some inclination from our point of view. So, we need to evaluate how this inclination affects the Q values we have measured. One way to avoid this problem is to correct the original observations by the tabulated inclination from previous studies, but this would not work, for example, if the disk is warped or has some morphology issues resulting from tidal force from a neighbor, or if the inclination is not well measured, etc.   

To understand  how the observed Q is or is not affected by inclination, we made models of the YSGs using the algorithm described in \cite{2004MNRAS.348..589C} for a determined fractal dimension, and we then calculated Q with the same procedure we have used for the real data.  We made one hundred models for each of these fractal dimensions: 1.4, 1.6, and 1.8, which are the typical values for the young stellar population (see Sect.~\ref{fractal_analysis}), and from inclination angles from 0 degrees to 85 degrees using a step of five degrees. For each of these dimensions and angles, we calculated the average Q. From the modeling we found that Q is not affected by inclination unless the angle is larger than 65 degrees. This high inclination angle makes Q lower than the value expected for its fractal dimension.
The inclination angle of the galaxy NGC~253 is $i=72$ degrees \citep{1991AJ....101..456P}. Thus, we de-projected the stars' positions in order to evaluate the inclination effects in our analysis. We found that on average the Q values vary very little, only by $\sim$0.05, and this is even smaller for NGC~2403 and NGC~300, for which the inclinations are lower than 65 degrees. 
Additionally, the values of Q in the main plane of the galaxy tend to be smaller, so the populations would be even more clustered.

In Fig.~\ref{mst}, we show some of the MST obtained in the three galaxies with different values of $Q$. In the first row, we show one of the YSGs with $Q \textgreater 0.8$ for each galaxy. We can see that the stars in these groups show a relatively homogeneous distribution without inner clumpings, unlike what is observed for the other YSGs with $Q \textless 0.8$.  It is also possible to appreciate that the smaller the value of $Q,$ the more clumps the YSG seems to present. 

In the top panels of Fig.~\ref{FigQ}, we plot $\overline{m}$ -- $\overline{s}$ behavior for the YSGs located on each galaxy, together with several constant $Q$ lines.
In the bottom panels we show the $Q$ distribution for the three galaxies, highlighting the $Q = 0.8$ and the mean values.
We observed that these distributions are similar in the three galaxies, with mean and median values of $\sim$0.65 in all of them. We also noticed the same trend for YSGs in NGC~247 \citep{2019A&A...626A..35R}, with a $Q_{mean}\sim 0.65$.

\section{The fractal behavior of the young population on a large scale}
\label{sect_fractal}

In Sect. \ref{MSTandQ}, we studied the internal structure of the YSGs in order to observe the hierarchical organization inside them. In this section, we study the distribution of young stars on a large scale. 

\subsection{Density maps}

For this task, we used the same selection of young blue stars used for the search of YSGs, \citep[see Sect. \ref{data} for NGC~2403, and][for NGC~253 and NGC~300, respectively]{2018MNRAS.479..961R,2016A&A...594A..34R}, to construct the density maps of these young populations. We convolved the stellar distribution with a Gaussian kernel using the kernel density estimation (KDE) method \citep[see][for details]{2019A&A...626A..35R}. 

\begin{figure*}
    \centering
    \includegraphics[trim={0 0 0 0},width=0.9\textwidth]{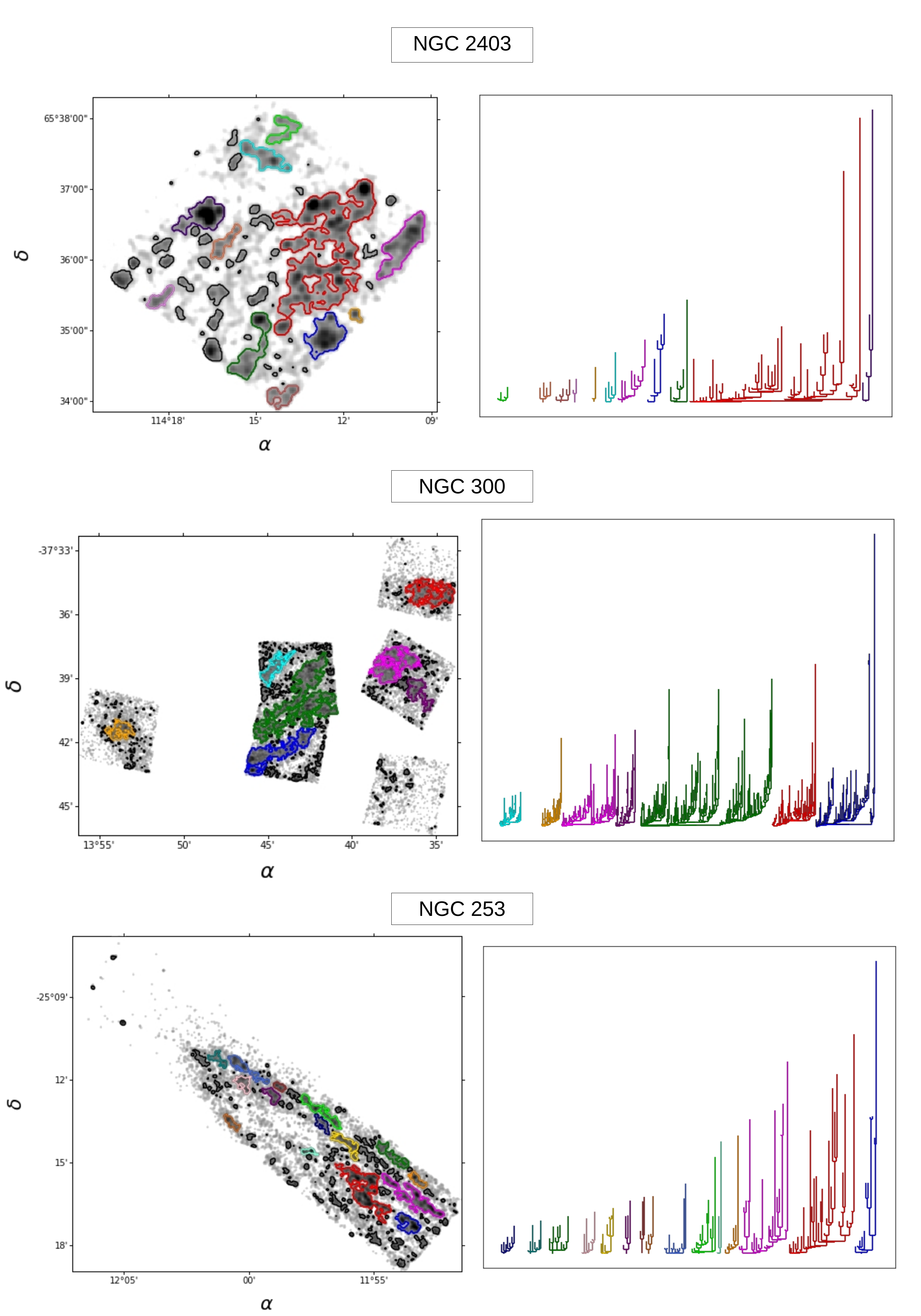}
    \caption{Left panels show density maps with their most prominent structures highlighted in color. On the right we present the dendrograms for these structures.}
    \label{maps}
\end{figure*}

\begin{figure*}
\centering
\resizebox{\hsize}{!}{\includegraphics{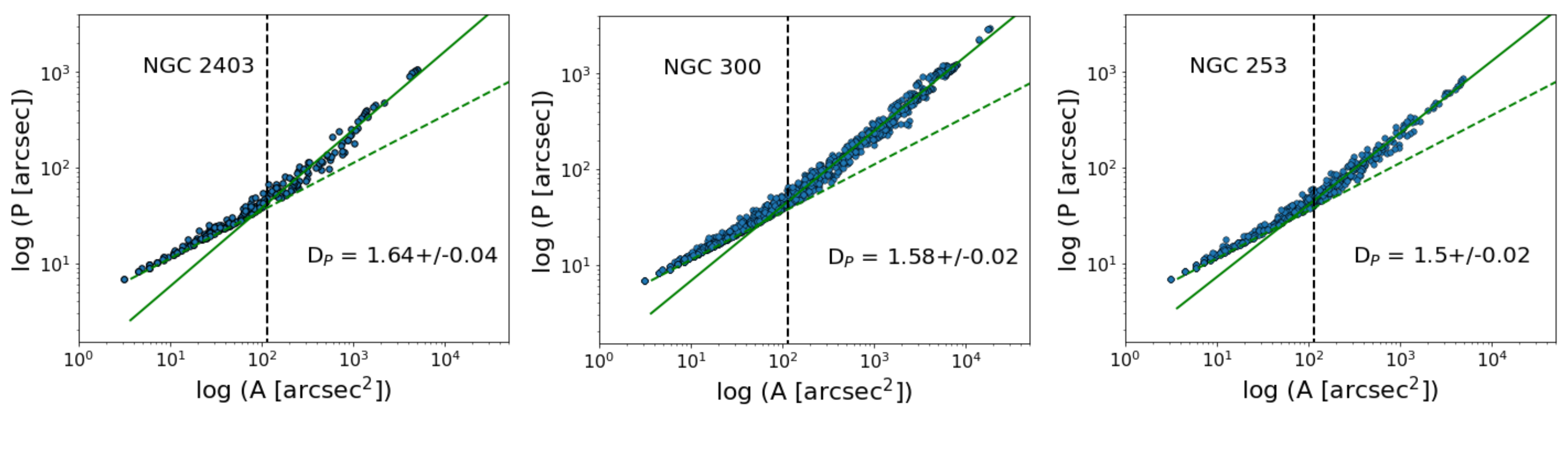}}
\caption{Perimeter-area relation for structures in density maps for the three galaxies. The vertical black line indicates the limit resolution of the map (see text), the green dashed line corresponds to the perimeter-area relation for a circle, the perimeter-area fit for the structures is shown with the continuous green line. The figure also shows the $D_P$ obtained in each case.}
\label{fractal}
\end{figure*}

\begin{figure}
\includegraphics[trim={0 20 0 20},width=\columnwidth]{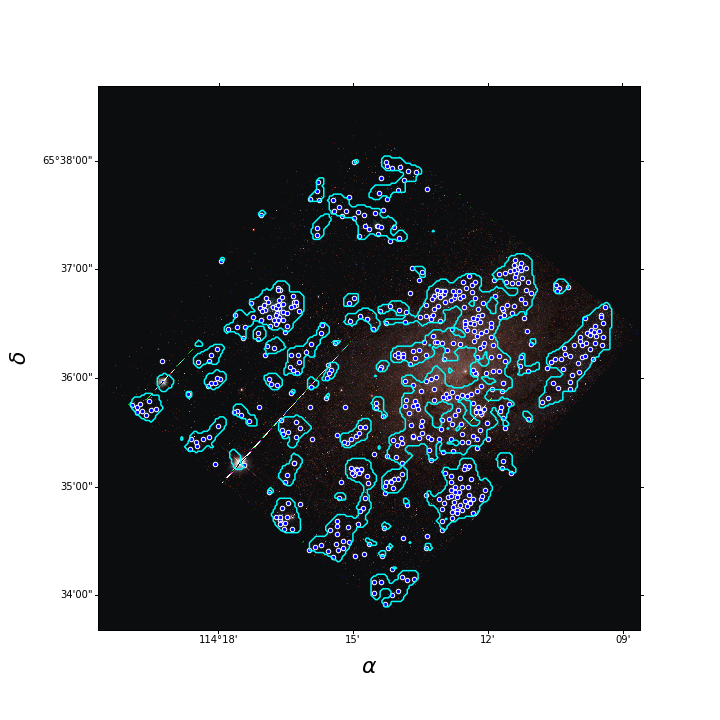}
\caption{Color image of observed region in NGC~2403. The contours of the structures detected in the density map are shown in light blue. The blue circles are the YSGs detected using the PLC method. We can observe an excellent correlation between YSGs and larger structures.}
\label{contours_2403}
\end{figure}

\begin{table*}[]
\centering
\caption{Summary of obtained parameters for the studied galaxies}
\begin{tabular}{lccccl}
\hline
Galaxy & $r$ [pc] & $\Gamma_{LF}$ & $Q$ & $D_P$ & Reference \\
\hline
NGC~247  & 60 & $\sim$ 0.2  & 0.65 $\pm$ 0.08 & 1.58 $\pm$ 0.02 & (1) \\
NGC~253  & 47 & $\sim$ 0.21 & 0.65 $\pm$ 0.08 & 1.50 $\pm$ 0.02 & (2), This work \\
NGC~300  & 50 & $\sim$ 0.13 & 0.64 $\pm$ 0.07 & 1.58 $\pm$ 0.02 & (3), This work \\
NCC~2403 & 45 & $\sim$ 0.16 & 0.65 $\pm$ 0.07 & 1.64 $\pm$ 0.04 & This work \\
\hline
\end{tabular}
\begin{minipage}{0.7\linewidth}
{\bf Notes:} (1) \cite{2019A&A...626A..35R}; (2) \cite{2018MNRAS.479..961R}; (3) \cite{2016A&A...594A..34R}
\end{minipage}
\label{tab:parameters}
\end{table*}

We adopted bandwidths of 2$",$ which correspond to $\sim$31~pc for NGC~2403, $\sim$ 34~pc for NGC~253, and $\sim$ 19~pc for NGC~300, respectively. 
This optimal bandwidth was selected after several tests with different values, we finally adopted the one that resulted in clear density maps with the main structures of the galaxies preserved. The obtained maps are shown in Fig.~\ref{maps} (right panel). We used these maps to find structures on various scales and at differing density levels. We did this using the "astrodendro" library of Python. In this figure, we can see the most prominent structures of each galaxy highlighted in color. In the left panels, we see the dendrograms corresponding to these structures (in the same color). The dendrograms  
show how a structure branches out into smaller and denser ones. The height of the branch indicates the density of the structure, with the highest branches being the densest ones.

These structures reach sizes of $\sim~760~pc$ in NGC~2403, $\sim~1064~pc$ in NGC~300, and $ \sim~730~pc$ in NGC~253.
The sizes were estimated using the statistic "radius" of the "astrodendro" library. It estimates the radius as the mean between the major and minor axis of the structures in the direction of the greatest elongation and its perpendicular, respectively. 
It should be noted that the structures in NGC~300 are limited in extension by the ACS/WFC 
field, and it is possible for them to reach larger sizes.
We note that the sizes of the structures may be affected by the inclination of the galaxy. The given values could be increased with respect to the real ones. In the highest inclination galaxy (NGC~253), the sizes could be up to 30\% smaller.

The obtained values are comparable with the maximum size for star forming structures obtained in the literature. \cite{1978PAZh....4..125E} found a characteristic size of 600~pc for stellar complexes in the Galaxy.
\cite{2010ApJ...720..541S} found a size scale between 500 and 1000~pc
as the limit for the self-similar structures generated by turbulent motions.
\cite{2017ApJ...842...25G} found a range between $\sim$200~pc to 1000~pc to the maximum size over which star formation is physically correlated.

In Fig.~\ref{contours_2403}, we show the color image of the studied region in NGC~2403, together with the contours of the main branches of the detected structures and the YSGs described in Sect.~\ref{PLC}. In this figure, it is possible to observe the excellent correlation between YSGs and structures, highlighting the hierarchical distribution. There are very few YSGs ($\sim 2 \%$) that fall outside the contours. The small percentage of YSGs that seem not to follow the hierarchical distribution are also found in NGC~247 and other galaxies \citep{2019A&A...626A..35R}.

\subsection{Fractal analysis}
\label{fractal_analysis}

\cite{mandelbrot1982fractal} proposed the fractal dimension ($D$) as a parameter to measure the degree of irregularity of an object. For an object in the space with a homogeneous distribution, the fractal dimension is 3 ($D_{3}$=3), and $D_{2}$=2 in a 2D space. If the object has irregularities $D_{3}<3$ or $D_{2}<2$, depending on the spatial dimensions. This parameter will take smaller values as the object becomes more irregular. The 2D fractal dimension can be estimated by the perimeter-area relation: $P \propto A^{D_P/2}$ \citep{mandelbrot1982fractal}. The value of $D_P$ expresses the way in which the contour of an object fills the space. For smooth shapes like a circle $D_P$=1 ($P \propto A^{1/2}$), if the contour of an object is very complicated, it will tend to fill all the space, in this case  $D_P$=2 ($P \propto A$) \citep{2010LNEA....4....1B}.

In Fig.~\ref{fractal}, we present the perimeter-area relation for the structures detected in the density maps for the three galaxies. The dashed green line shows the perimeter-area relation for a circle, and the vertical black line indicates the resolution limit of the maps ($3 \times bandwidth$). We noticed that structures below this limit follow the perimeter-area relation for a circle. This is because in this region we do not have enough resolution to see the real shape of the contours. We fit the $D_P$ for structures above this limit (continuous green line), obtaining the values indicated in Fig.~\ref{fractal} and presented in Table~\ref{tab:parameters}.

Our resulting estimated $D_P$ values are consistent with fractal distributions and also are in good agreement with those obtained in other galaxies (see Table~\ref{tab:parameters}) for young stellar structures and the ISM. 
\cite{2017ApJ...835..171S} found a fractal dimension $D_2=1.6\pm0.03$ for the stellar structures in the 30 Doradus regions in the LMC, and \citet{2017ApJ...849..149S} a value of $D_2=1.5\pm0.1$ for structures in the LMC's bar complex. For the young stellar structures in the Small Magellanic Cloud (SMC), \cite{2018ApJ...858...31S} found a projected fractal dimension of $1.44 \pm 0.02$. \cite{2009A&A...494...81S} found a value of $D_P$ between 1.2 and 1.6 for young star clusters in M~51. \cite{2008ApJS..178....1S} studied the HII regions in a large sample of disk galaxies, finding an average value of 1.81 for the 2D fractal dimension.
The projected fractal dimension in the ISM has typical values of $\sim 1.3 - 1.5$ \citep[e.g.,][]{2004JKAS...37..137L,2007ARA&A..45..339B}. 
We discuss these results in the following section.

\section{Discussion}

\label{discuss}

The structures in the ISM present a fractal or hierarchical distribution, observed from the GMCs to clumps and dense cores, in a size range of $\sim 0.01 - 1000~pc$. This distribution is shaped by turbulent motions \citep{2004ARA&A..42..211E}, with a characteristic projected fractal dimension of 1.3-1.4 \citep{2007ARA&A..45..339B,2010LNEA....4....1B}. The projected fractal dimension in 2D for fractal objects has values between 1 and 2 \citep{Florio2019}. In this work, we found fractal dimensions consistent with a fractal distribution for three nearby galaxies, strengthening the idea that the observed fractal behavior of young stars is passed on from the structures in the ISM from which they form. 

In a hierarchical star formation model regulated by turbulence and self-gravity, the  
gas compression in the GMC fragmented the cloud into successive smaller clouds, creating the observed hierarchical distribution in the ISM \citep{1999ApJ...527..266E, 2017ApJ...842...25G}. The densest region of this distribution will form stars. Thus, 
the fractal behavior would be inherited by the recently born stars. Therefore, their initial space distribution is expected to be related with the density peaks in the GMCs. 

The young population of NGC~2403 studied in this work features a range of ages between $\sim$ 10 and 90 Myr. By means of the $Q$ parameter (see Sect. \ref{MSTandQ}), we find that $\sim 98\%$ of the groups still preserve the internal hierarchical distribution of their parent molecular clouds. This means that the timescale in which the original distribution of stars is erased is larger than 90~Myr for this galaxy. For NGC~300, this time should be $>$ 235 Myr and $>$ 75 Myr for NGC~253.

\section{Conclusions}

\label{conclusions}

We studied the hierarchical organization and fractal behavior of the young stellar population in three nearby galaxies. We first detected 573 YSGs in NGC~2403 by means of the PLC method. For these YSGs, we determine a mean size of 45~pc, which is similar to the size found in NGC~300, NGC~253, and other nearby galaxies, and it is slightly lower than that of the YSGs in NGC~247. We also find a mean LF slope of 0.16 and a mean density of $\sim0.001~stars.pc^{-3}$ counting stars brighter than $B7$. We built a catalog with the main properties of each detected YSG, which is available on the CDS.

Then, we studied the internal distribution of the 573 YSGs in NGC~2403 mentioned before, plus 1147 YSGs in NGC~300, and the 875 YSGs in NGC~253 detected in previous works. We find that most of the YSGs in all these galaxies present values of $Q < 0.8$, which means that the YSGs have inner clumpings. The same trend was also found in our previous work for NGC~247. This fact would indicate that the studied YSGs are very young and still retain the distribution of the molecular clouds from which they formed.

We also identified young stellar structures on larger scales through the stellar density maps. We observed, by means of the dendrograms, that this population presents a high degree of clustering. In the three galaxies, we can see large structures that split themselves into smaller and denser structures over several densities levels. We derived projected fractal dimensions for these structures. 
These values indicate that the young population in these galaxies presents a fractal behavior. They are also similar to the ones found for the young population in other galaxies and in the ISM, suggesting that the young stellar structures obtain this distribution from their parent molecular clouds, reinforcing the idea of star formation driven by turbulence and self-gravity.


\bibliographystyle{aa} 
\bibliography{biblio} 

\begin{thebibliography}{48}
\expandafter\ifx\csname natexlab\endcsname\relax\def\natexlab#1{#1}\fi

\bibitem[{{Ann} {et~al.}(2015){Ann}, {Seo}, \& {Ha}}]{2015ApJS..217...27A}
{Ann}, H.~B., {Seo}, M., \& {Ha}, D.~K. 2015, ApJS, 217, 27

\bibitem[{{Battinelli}(1991)}]{1991A&A...244...69B}
{Battinelli}, P. 1991, \aap, 348, 589

\bibitem[{{Bergin} \& {Tafalla}(2007)}]{2007ARA&A..45..339B}
{Bergin}, E.~A. \& {Tafalla}, M. 2007, \araa, 45, 339

\bibitem[{{Blaauw}(1964)}]{1964ARA&A...2..213B}
{Blaauw}, A. 1964, \araa, 2, 213

\bibitem[{{Caballero}(2008)}]{2008MNRAS.383..375C}
{Caballero}, J.~A. 2008, \mnras, 383, 375

\bibitem[{{Cartwright} \& {Whitworth}(2004)}]{2004MNRAS.348..589C}
{Cartwright}, A. \& {Whitworth}, A.~P. 2004, \mnras, 348, 589

\bibitem[{{Clarke} {et~al.}(2000){Clarke}, {Bonnell}, \&
  {Hillenbrand}}]{2000prpl.conf..151C}
{Clarke}, C.~J., {Bonnell}, I.~A., \& {Hillenbrand}, L.~A. 2000, in Protostars
  and Planets IV, ed. V.~{Mannings}, A.~P. {Boss}, \& S.~S. {Russell}, 151

\bibitem[{{Dalcanton} {et~al.}(2009){Dalcanton}, {Williams}, {Seth}, {Dolphin},
  {Holtzman}, {Rosema}, {Skillman}, {Cole}, {Girardi}, {Gogarten},
  {Karachentsev}, {Olsen}, {Weisz}, {Christensen}, {Freeman}, {Gilbert},
  {Gallart}, {Harris}, {Hodge}, {de Jong}, {Karachentseva}, {Mateo}, {Stetson},
  {Tavarez}, {Zaritsky}, {Governato}, \& {Quinn}}]{2009ApJS..183...67D}
{Dalcanton}, J.~J., {Williams}, B.~F., {Seth}, A.~C., {et~al.} 2009, \apjs,
  183, 67

\bibitem[{{Drissen} {et~al.}(1999){Drissen}, {Roy}, {Moffat}, \&
  {Shara}}]{1999AJ....117.1249D}
{Drissen}, L., {Roy}, J.-R., {Moffat}, A.~F.~J., \& {Shara}, M.~M. 1999, AJ,
  117, 1249

\bibitem[{{Efremov}(1978)}]{1978PAZh....4..125E}
{Efremov}, Y.~N. 1978, Pisma v Astronomicheskii Zhurnal, 4, 125

\bibitem[{{Elmegreen}(1993)}]{1993ApJ...419L..29E}
{Elmegreen}, B.~G. 1993, \apjl, 419, L29

\bibitem[{{Elmegreen}(1999)}]{1999ApJ...527..266E}
{Elmegreen}, B.~G. 1999, \apj, 527, 266

\bibitem[{{Elmegreen} \& {Efremov}(1996)}]{1996ApJ...466..802E}
{Elmegreen}, B.~G. \& {Efremov}, Y.~N. 1996, \apj, 466, 802

\bibitem[{{Elmegreen} \& {Falgarone}(1996)}]{1996ApJ...471..816E}
{Elmegreen}, B.~G. \& {Falgarone}, E. 1996, \apj, 471, 816

\bibitem[{{Elmegreen} \& {Scalo}(2004)}]{2004ARA&A..42..211E}
{Elmegreen}, B.~G. \& {Scalo}, J. 2004, \araa, 42, 211

\bibitem[{{Elmegreen} \& {Salzer}(1999)}]{1999AJ....117..764E}
{Elmegreen}, D.~M. \& {Salzer}, J.~J. 1999, \aj, 117, 764

\bibitem[{{Fernandes} {et~al.}(2012){Fernandes}, {Gregorio-Hetem}, \&
  {Hetem}}]{2012A&A...541A..95F}
{Fernandes}, B., {Gregorio-Hetem}, J., \& {Hetem}, A. 2012, \aap, 541, A95

\bibitem[{Florio {et~al.}(2019)Florio, Fawell, \& Small}]{Florio2019}
Florio, B., Fawell, P., \& Small, M. 2019, Powder Technology, 343, 551

\bibitem[{{Gouliermis}(2018)}]{2018PASP..130g2001G}
{Gouliermis}, D.~A. 2018, \pasp, 130, 072001

\bibitem[{{Gouliermis} {et~al.}(2017){Gouliermis}, {Elmegreen}, {Elmegreen},
  {Calzetti}, {Cignoni}, {Gallagher}, {Kennicutt}, {Klessen}, {Sabbi},
  {Thilker}, {Ubeda}, {Aloisi}, {Adamo}, {Cook}, {Dale}, {Grasha}, {Grebel},
  {Johnson}, {Sacchi}, {Shabani}, {Smith}, \& {Wofford}}]{2017MNRAS.468..509G}
{Gouliermis}, D.~A., {Elmegreen}, B.~G., {Elmegreen}, D.~M., {et~al.} 2017,
  \mnras, 468, 509

\bibitem[{{Gouliermis} {et~al.}(2015){Gouliermis}, {Thilker}, {Elmegreen},
  {Elmegreen}, {Calzetti}, {Lee}, {Adamo}, {Aloisi}, {Cignoni}, {Cook}, {Dale},
  {Gallagher}, {Grasha}, {Grebel}, {Dav{\'o}}, {Hunter}, {Johnson}, {Kim},
  {Nair}, {Nota}, {Pellerin}, {Ryon}, {Sabbi}, {Sacchi}, {Smith}, {Tosi},
  {Ubeda}, \& {Whitmore}}]{2015MNRAS.452.3508G}
{Gouliermis}, D.~A., {Thilker}, D., {Elmegreen}, B.~G., {et~al.} 2015, \mnras,
  452, 3508

\bibitem[{{Grasha} {et~al.}(2017){Grasha}, {Elmegreen}, {Calzetti}, {Adamo},
  {Aloisi}, {Bright}, {Cook}, {Dale}, {Fumagalli}, {Gallagher}, {Gouliermis},
  {Grebel}, {Kahre}, {Kim}, {Krumholz}, {Lee}, {Messa}, {Ryon}, \&
  {Ubeda}}]{2017ApJ...842...25G}
{Grasha}, K., {Elmegreen}, B.~G., {Calzetti}, D., {et~al.} 2017, \apj, 842, 25

\bibitem[{{Gregorio-Hetem} {et~al.}(2015){Gregorio-Hetem}, {Hetem},
  {Santos-Silva}, \& {Fernand es}}]{2015MNRAS.448.2504G}
{Gregorio-Hetem}, J., {Hetem}, A., {Santos-Silva}, T., \& {Fernand es}, B.
  2015, \mnras, 448, 2504

\bibitem[{{Hodge}(1985)}]{1985PASP...97.1065H}
{Hodge}, P. 1985, PASP, 97, 1065

\bibitem[{{Krause} {et~al.}(2020){Krause}, {Offner}, {Charbonnel}, {Gieles},
  {Klessen}, {V{\'a}zquez-Semadeni}, {Ballesteros-Paredes}, {Girichidis},
  {Kruijssen}, {Ward}, \& {Zinnecker}}]{2020SSRv..216...64K}
{Krause}, M. G.~H., {Offner}, S. S.~R., {Charbonnel}, C., {et~al.} 2020, \ssr,
  216, 64

\bibitem[{{Lada} \& {Lada}(2003)}]{2003ARA&A..41...57L}
{Lada}, C.~J. \& {Lada}, E.~A. 2003, \araa, 41, 57

\bibitem[{{Larson}(2003)}]{2003RPPh...66.1651L}
{Larson}, R.~B. 2003, Reports on Progress in Physics, 66, 1651

\bibitem[{{Lee}(2004)}]{2004JKAS...37..137L}
{Lee}, Y. 2004, Journal of Korean Astronomical Society, 37, 137

\bibitem[{Mandelbrot(1982)}]{mandelbrot1982fractal}
Mandelbrot, B.~B. 1982, The fractal geometry of nature, Vol.~1 (WH freeman New
  York)

\bibitem[{{McKee} \& {Ostriker}(2007)}]{2007ARA&A..45..565M}
{McKee}, C.~F. \& {Ostriker}, E.~C. 2007, \araa, 45, 565

\bibitem[{{Puche} {et~al.}(1991){Puche}, {Carignan}, \& {van
  Gorkom}}]{1991AJ....101..456P}
{Puche}, D., {Carignan}, C., \& {van Gorkom}, J.~H. 1991, \aj, 101, 456

\bibitem[{{Rodr{\'\i}guez} {et~al.}(2016){Rodr{\'\i}guez}, {Baume}, \&
  {Feinstein}}]{2016A&A...594A..34R}
{Rodr{\'\i}guez}, M.~J., {Baume}, G., \& {Feinstein}, C. 2016, \aap, 594, A34

\bibitem[{{Rodr{\'\i}guez} {et~al.}(2018){Rodr{\'\i}guez}, {Baume}, \&
  {Feinstein}}]{2018MNRAS.479..961R}
{Rodr{\'\i}guez}, M.~J., {Baume}, G., \& {Feinstein}, C. 2018, \mnras, 479, 961

\bibitem[{{Rodr{\'\i}guez} {et~al.}(2019){Rodr{\'\i}guez}, {Baume}, \&
  {Feinstein}}]{2019A&A...626A..35R}
{Rodr{\'\i}guez}, M.~J., {Baume}, G., \& {Feinstein}, C. 2019, \aap, 626, A35

\bibitem[{{S{\'a}nchez} {et~al.}(2010){S{\'a}nchez}, {A{\~n}ez}, {Alfaro}, \&
  {Crone Odekon}}]{2010ApJ...720..541S}
{S{\'a}nchez}, N., {A{\~n}ez}, N., {Alfaro}, E.~J., \& {Crone Odekon}, M. 2010,
  \apj, 720, 541

\bibitem[{{S{\'a}nchez} \& {Alfaro}(2008)}]{2008ApJS..178....1S}
{S{\'a}nchez}, N. \& {Alfaro}, E.~J. 2008, \apjs, 178, 1

\bibitem[{{S{\'a}nchez} \& {Alfaro}(2009)}]{2009ApJ...696.2086S}
{S{\'a}nchez}, N. \& {Alfaro}, E.~J. 2009, \apj, 696, 2086

\bibitem[{{S{\'a}nchez} \& {Alfaro}(2010)}]{2010LNEA....4....1B}
{S{\'a}nchez}, N. \& {Alfaro}, E.~J. 2010, {The fractal spatial distribution of
  stars in open clusters and stellar associations}, Vol.~4, 1--11

\bibitem[{{Scheepmaker} {et~al.}(2009){Scheepmaker}, {Lamers}, {Anders}, \&
  {Larsen}}]{2009A&A...494...81S}
{Scheepmaker}, R.~A., {Lamers}, H.~J.~G.~L.~M., {Anders}, P., \& {Larsen},
  S.~S. 2009, \aap, 494, 81

\bibitem[{{Schlafly} \& {Finkbeiner}(2011)}]{2011ApJ...737..103S}
{Schlafly}, E.~F. \& {Finkbeiner}, D.~P. 2011, \apj, 737, 103

\bibitem[{{Schmeja} \& {Klessen}(2006)}]{2006A&A...449..151S}
{Schmeja}, S. \& {Klessen}, R.~S. 2006, \aap, 449, 151

\bibitem[{{Sivan} {et~al.}(1990){Sivan}, {Petit}, {Comte}, \&
  {Maucherat}}]{1990A&A...237...23S}
{Sivan}, J.~P., {Petit}, H., {Comte}, G., \& {Maucherat}, A.~J. 1990, \aap,
  237, 23

\bibitem[{{Sun} {et~al.}(2018){Sun}, {de Grijs}, {Cioni}, {Rubele},
  {Subramanian}, {van Loon}, {Bekki}, {Bell}, {Ivanov}, {Marconi}, {Muraveva},
  {Oliveira}, \& {Ripepi}}]{2018ApJ...858...31S}
{Sun}, N.-C., {de Grijs}, R., {Cioni}, M.-R.~L., {et~al.} 2018, \apj, 858, 31

\bibitem[{{Sun} {et~al.}(2017{\natexlab{a}}){Sun}, {de Grijs}, {Subramanian},
  {Bekki}, {Bell}, {Cioni}, {Ivanov}, {Marconi}, {Oliveira}, {Piatti},
  {Ripepi}, {Rubele}, {Tatton}, \& {van Loon}}]{2017ApJ...849..149S}
{Sun}, N.-C., {de Grijs}, R., {Subramanian}, S., {et~al.} 2017{\natexlab{a}},
  \apj, 849, 149

\bibitem[{{Sun} {et~al.}(2017{\natexlab{b}}){Sun}, {de Grijs}, {Subramanian},
  {Cioni}, {Rubele}, {Bekki}, {Ivanov}, {Piatti}, \&
  {Ripepi}}]{2017ApJ...835..171S}
{Sun}, N.-C., {de Grijs}, R., {Subramanian}, S., {et~al.} 2017{\natexlab{b}},
  \apj, 835, 171

\bibitem[{{Tang} {et~al.}(2014){Tang}, {Bressan}, {Rosenfield}, {Slemer},
  {Marigo}, {Girardi}, \& {Bianchi}}]{2014MNRAS.445.4287T}
{Tang}, J., {Bressan}, A., {Rosenfield}, P., {et~al.} 2014, \mnras, 445, 4287

\bibitem[{{Tully} {et~al.}(2013){Tully}, {Courtois}, {Dolphin}, {Fisher},
  {H{\'e}raudeau}, {Jacobs}, {Karachentsev}, {Makarov}, {Makarova},
  {Mitronova}, {Rizzi}, {Shaya}, {Sorce}, \& {Wu}}]{2013AJ....146...86T}
{Tully}, R.~B., {Courtois}, H.~M., {Dolphin}, A.~E., {et~al.} 2013, AJ, 146, 86

\bibitem[{{V{\'a}zquez-Semadeni} {et~al.}(2017){V{\'a}zquez-Semadeni},
  {Gonz{\'a}lez-Samaniego}, \& {Col{\'\i}n}}]{2017MNRAS.467.1313V}
{V{\'a}zquez-Semadeni}, E., {Gonz{\'a}lez-Samaniego}, A., \& {Col{\'\i}n}, P.
  2017, \mnras, 467, 1313

\end{thebibliography}

\begin{acknowledgements}
We thank the referee for helpful comments and constructive suggestions that helped to improve this paper.
MJR and GB acknowledge support from CONICET (PIP 112-201701-00055). MJR is a fellow of CONICET. This work was based on observations made with the NASA/ESA Hubble Space Telescope, and obtained from the Hubble Legacy Archive, which is a collaboration between the Space Telescope Science Institute (STScI/NASA), the Space Telescope European Coordinating Facility (ST-ECF/ESA), and the Canadian Astronomy Data Centre (CADC/NRC/CSA). Some of the data presented in this paper were obtained from the Mikulski Archive for Space Telescopes (MAST). STScI is operated by the Association of Universities for Research in Astronomy, Inc., under NASA contract NAS5-26555. Support for MAST for non-HST data is provided by the NASA Office of Space Science via grant NNX09AF08G and by other grants and contracts. This research has made use of "Aladin sky atlas" developed at CDS, Strasbourg Observatory, France. We also made use of astrodendro, a Python package used to compute dendrograms of astronomical data (http://www.dendrograms.org/).
     
\end{acknowledgements}

\end{document}